# Experimental realization of a carpet cloak for temperature field and heat flux

Tianzhi Yang[1]*,   Weikai Xu[1],   Lujun Huang[2],   Xiaodong Yang[3],   Fei Chen[1]

1. School of Aerospace Engineering, Shenyang Aerospace University, Shenyang 110136, China

2. National Laboratory for Infrared Physics, Shanghai Institute of Technical Physics, Shanghai, 200083, China

3. College of Mechanical Engineering, Beijing University of Technology, Beijing 100124, China

**Abstract**

Based on transformation optics (TO), we present and experimentally realize a new thermal carpet cloak. The device, which we call a "thermal carpet", provides a considerable cloaking effect. Moreover, the device can maximally suppress the heat flux distortion around an object. The device is designed, fabricated and measured to verify the thermal cloaking performance. In comparison with previous experimental work, the advantage of this design is that the required medium parameter is inherently isotropic.
.

The transformation optics (TO) has been proposed as a powerful tool to control over wave propagation. The concept was initially proposed in the context of electromagnetics, and later extended to acoustics [1-6]. One of the most important applications of TO is "cloak", a coating shell or slab can guide the propagation of light and acoustic waves. As a result, a region inside the shell becomes invisible.

Later, TO inspired many theoretical and experimental approaches for controlling other waves, such as dc magnetic field [7], elastic wave [8], quantum [9] and matter waves [10]. Very recently, the TO theory was also extended to thermodynamics [11,12]. The heat flux transfer is quite

---

[1]* Corresponding author

*Email addres*s: yangtz@me.com

different from wave behavior, because it does not transport energy as normally waves do. However, due to the invariance of heat conduction equation, the thermal cloak by using TO was still achieved recently. This is because the heat conduction equation, also known as the heat diffusion equation, is invariance under coordinate transformation. The pioneer works open possibilities for cloaking and focusing heat flux. Thus, the transformation thermodynamics may provide a new strategy to manipulate heat flux or channel thermal energy.

On the other hand, such above mentioned heat flux cloaks generally require spatially varying constitutive parameters, results in extremely complex material parameters. Thus, such devices are very inhomogeneous and very difficult to fabricate. Such as the required thermal conductivities inside the device are close to zero or vary large. Due to such extreme properties, only a few experimental works of thermal cloaks exist so far [13-17].

In this letter, we present and fabricate the ground cloak labeled "thermal carpet" or "thermal ground cloak" to shield a triangle region from diffusive heat flow. The artificial device is designed with a multilayered structure with inherent isotropic materials. Such device is expected to protect electronics, microprocessor, micro-combustion and batteries.

The material parameters of the thermal carpet will be derived based on the TO theory. Here we start from the thermal conduction equation without the source term

$$\nabla(-\kappa\nabla T)=0 \qquad (1)$$

where $\kappa$ is the thermal conductivity and $T$ is the temperature. In this study, the analysis is restricted to a two dimensional case.

Figure 1 shows the principle of the thermal carpet. The shaded region is the designed artificial composite and the bottom triangular region is the region to be cloaked. To obtain a ground plane cloak, we consider the following transformations [18-20]

$$\begin{aligned} x' &= x \\ y' &= \frac{c-a}{c}y + \frac{b-x\,\mathrm{sgn}(x)}{b}a \\ z' &= z \end{aligned} \qquad (2)$$

where $a$, $b$ and $c$ are the geometric parameters.

Note that Eq. (2) is a linear transformation. According to the TO theory, the effective thermal conductivity can be derived by using the following equation

$$\tilde{\kappa}(x') = \frac{A\kappa(x)A^T}{\det(A)} \qquad (3)$$

where $A$ is the Jacobian transformation matrix with elements defined by

$$A = \frac{\partial x'_i}{\partial x_i} \qquad i=1, 2, 3 \qquad (4)$$

By using the mapping functions (2)-(4), the heat conduction Eq. (1) is mapped into:

$$\nabla(-\tilde{\kappa}\nabla T)=0 \qquad (5)$$

where the transformed thermal conductivity tensor $\tilde{\kappa}$ can be expressed as

$$\tilde{\kappa} = \begin{bmatrix} \dfrac{c}{c-a} & -\dfrac{ac}{(c-a)b}\mathrm{sgn}(x) \\ -\dfrac{ac}{(c-a)b}\mathrm{sgn}(x) & \dfrac{c-a}{c} + \dfrac{c}{c-a}\left(\dfrac{a}{b}\right)^2 \end{bmatrix} \qquad (6)$$

Eq. (6) is the expression of the required physical effective material parameters of the thermal carpet.

Numerical simulations of the thermal cloak are carried out by means of COMSOL Multiphysics. In Fig. 2 (a) and (b), we show that the thermal carpet can shield the bottom triangle region incoming heat flow from both top and left sides. The temperature inside the cloaked region is much lower than external values. Thus, such device could be used to protect a certain region when a temperature gradient is suddenly imposed and minimized the temperature variation in the cloaked region.

It is noted that the anisotropic thermal conductivity needed in Eq. (6) could be obtained in the principal axis system, in which the components of tensor $\tilde{\kappa}$ are

$$\tilde{\kappa} = \kappa_0 \begin{bmatrix} \tilde{\kappa}_{pr}^{(11)} & 0 \\ 0 & \tilde{\kappa}_{pr}^{(22)} \end{bmatrix} = \kappa_0 \begin{bmatrix} F+\sqrt{F^2-1} & 0 \\ 0 & F-\sqrt{F^2-1} \end{bmatrix} \qquad (7)$$

where $\tilde{\kappa}_{pr}^{(11)}$ and $\tilde{\kappa}_{pr}^{(22)}$ are the in-plane diagonal components of the principal axis thermal conductivity and

$$F = \frac{b^2c^2 + (c-a)^2b^2 + a^2b^2}{2(c-a)b^2c}$$

The angle of rotation is

$$\alpha = \text{sgn}(x)\left\{\arctan[\frac{bc - b(c-a)(F + \sqrt{F^2 - 1})}{ac}] + \frac{\pi}{2}\right\} \qquad (8)$$

These two components satisfy a special relationship

$$\tilde{\kappa}_{pr}^{(11)}\tilde{\kappa}_{pr}^{(22)} = (F + \sqrt{F^2 - 1})(F - \sqrt{F^2 - 1}) = 1 \qquad (9)$$

Eq. (9) indicates the thermal cloak can be realized by using inherent homogeneous and isotropic material with finite thermal conductivity. It is noted that a similar relation is also found by Han et al. [21] for annular cloaking region.

Based on an experimental optimization, we choose $a$=70mm, $b$=71.2mm, $c$=93.3mm, the homogeneous material parameters can be evaluated in the principle system

$$\tilde{\kappa}_{pr} = \kappa_0 \begin{bmatrix} 8.0277 & 0 \\ 0 & 0.1246 \end{bmatrix} \qquad (10)$$

and the ideal rotating angle of the principal system is $\alpha \approx 44.67^o$. Although it is generally impossible to find a natural material exhibiting the desired constitutive parameters in Eq. (10), it still can be approximated by engineered multilayered composites. We can alternatively stack two sheets with isotropic thermal conductivity $\kappa_A$ and $\kappa_B$. In this paper, the required conductivity of the background is achieved by grooving on a stainless steel plate and filling the grooves with cork. The corresponding thermal conductivity of the stainless steel (0Cr18Ni9, ASME 304) are $\kappa_{steel}$ =16.2 W m$^{-1}$ K$^{-1}$ and $\kappa_{cork}$=0.065 W m$^{-1}$ K$^{-1}$, respectively. The cork area fraction $\eta$ is obtained according to the 2D effective media theory

$$\frac{1}{\tilde{\kappa}_{pr}^{(11)}} = \frac{1}{1+\eta}(\frac{1}{\kappa_A} + \frac{1}{\kappa_B}), \quad \tilde{\kappa}_{pr}^{(22)} = \frac{\kappa_A + \eta\kappa_B}{1+\eta} \qquad (11)$$

The thickness of the stainless plate is chosen as 2.5 mm to minimize the heat convection by the air. In Table 1, we compare the values of thermal conductivities between and desired and practical implement. It is seen that the relative error is quite small, indicating that such design can be realized by such layered materials.

Table 1 Comparison between desired and practical implement

| $\eta=1.02$ | Desired | Practical | Error |
|---|---|---|---|
| $\tilde{\kappa}_{pr}^{(11)}$ | 8.0277 | 8.0520 | 0.3% |
| $\tilde{\kappa}_{pr}^{(22)}$ | 0.1246 | 0.1282 | 2.93% |

The required physical material parameters throughout the thermal carpet are not extreme, leading to an easier way to realized in practice. The experiment setup and the fabricated carpet cloak device are shown in Figure 3 (a) and in Fig. 3 (b), repspectively. The required conductivity of the background is achieved by drilling holes into a stainless steel plate and filling them with polydimethylsiloxane (PDMS) with thermal conductivity of $\kappa_{PDMS}$=0.17 W m$^{-1}$ K$^{-1}$. The PDMS fraction $f$ can be obtained from the Maxwell-Garnett formula $\kappa_b = \kappa_{steel}[1 + \frac{2f(\kappa_{PDMS} - \kappa_{steel})}{\kappa_{PDMS} + \kappa_{steel} - f(\kappa_{PDMS} - \kappa_{steel})}]$. By performing an experimental optimization, we choose a square lattice with lattice constant 3mm and the holes with diameter 2.3mm, leading to $f$=46.16%. Thus we obtain $\kappa_b$=6.11 W m$^{-1}$ K$^{-1}$.

In the experimental setup, the local heating and thermal sink are achieved by using a controllable heater and ice-water mixture, respectively. To capture the temperature profile using conventional infrared camera, we need an additional coating of the stainless surface to achieve a nearly black body. Thus, to improve the thermal emission efficiency, we isolated both surfaces of the sample by an approximately 150 μm thin layer of blackbody paint. Figure 4 (a) and (b) show measured temperature distributions of the sample for different higher temperature at the top side. The top side of the sample is placed on a heater with a controllable higher temperature, and the bottom is placed on a tank with ice-water mixture, which has a lower temperature fixed at 0°C. Due to the temperature gradient, heat flux diffuses from top side (hot) toward the bottom side (cold). As a result, heat flux transport takes place until the system has become uniform temperature throughout. It is found that the average temperature inside the cloaked region is much lower than its surrounding.

More importantly, the isothermal lines on the sample are nearly horizontal, indicating a temperature distortion is minimized. As a result, the isothermal lines outside the cloaked region would not reveal the presence of the carpet. It is known that an ordinary composite with flux path will induce isothermal distortion. Our simple isotropic carpet can maximally suppress the heat flux distortion around an object.

As another case, the right side of the sample is placed on the heater and left side is placed on the ice-water tank, respectively. Thus heat flux flows from right to left. The stationary temperature profile inside the rectangular region is shown in Figure 4 (c) and (d), respectively. It is seen that the heat flux propagation is apparently re-directed the bottom triangular region is also cloaked. The external heat flux is significantly reduced to enter the cloaking region. Thus, such carpet could be used to protect a certain region when a temperature gradient is suddenly imposed and minimized the temperature variation in the cloaked region.

In conclusion, we experimentally realize a carpet thermodynamic cloak. In general, such carpet could be used to protect a certain region from the invasion of the eternal heat flux. More importantly, comparing with previous work, this flat carpet is easier to fabricate and avoid the singular problem of present in previous cloaking device. The cloaking performance is verified for different incoming directions of heat flux. Such simple design can maximally suppress the isothermal lines distortion and eliminated the disturbance around the carpet. We hope that such device could lead to practical applications, for instance, thermal management and thermal imaging.

## Acknowledgements

This work was supported by the National Natural Science Foundation of China (No. 11202140, 10702045, 11302135 and 11172010), the National Science Fund for Excellent Young Scholars (No.11322214), Program for Liaoning Excellent Talents in University (No. LJQ2013020) and State Key Laboratory of Mechanics and Control of Mechanical Structures (MCMS-0112G01).

## Reference

[1] J.B.Pendry, D.Schurig, and D.R.Smith, Science, **312**, 1780 (2006)
[2] U. Leonhardt, Science, **312**, 1777 (2006)
[3] G.W.Milton, M.Briane, and J.R.Willis, New J. Phys. **8**, 248 (2006)
[4] H.Chen, and C.T.Chan, Appl. Phys. Lett. **91**, 183518 (2006)
[5] A.N.Norris, Proc. Roy. Soc. A. **34**, 464 (2008)
[6] S.Zhang, C.Xia, and N.Fang, Phys. Rev. Lett. **106**, 024301 (2011)

[7] W.Zhu, C.Ding, and X.Zhao, Appl. Phys. Lett. **97**, 131902 (2010)

[8] M.Farhat, New J. Phys. **10**, 115030 (2008)

[9] A.Greenleaf, Y. Kurylev, M. Lassas, and G. Uhlmann, Phys. Rev. Lett. **101**, 220404 (2008).

[10] M.Brun, S. Guenneau, and A.B.Movchan, Appl. Phys. Lett. **94**, 061903 (2008)

[11] C. Z. Fan, Y. Gao, and J. P. Huang, Appl. Phys. Lett. **92**, 251907 (2008).

[12] S. Guenneau, C.Amra, and D.Veynante, Opt.Expr. **20**, 8202 (2012)

[13] T. Han, X. Bai, D. Gao, J. T. L. Thong, B. Li, and C.W. Qiu, Phys. Rev. Lett. **112**, 054302 (2014).

[14] S. Narayana and Y.Sato, Phys. Rev. Lett. **108**, 214303 (2012)

[15] R. Schittny, M. Kadic, S. Guenneau, and M. Wegener, Phys. Rev. Lett. **110**, 195901 (2013).

[16] H.Xu, X.Shi, F.Gao, H.Sun, and B.Zhang, Phys. Rev. Lett. **112**, 054301 (2014)

[17] S. Narayana and Y.Sato, Appl. Phys. Lett. **102**, 201904 (2013)

[18] J.Li, and J.B.Pendry, Phys. Rev. Lett. **101**, 203901 (2008).

[19] H.F.Ma, and T. J.Cui, Nature Comm. **1**, 1 (2010)

[20] B.Popa, L.Zigoneanul, and S.A.Cummer, Phys. Rev. Lett. **106**, 253901 (2011)

[21] T. Han, B. Li, and C.W.Qiu, Sci. Report, **3**, 1593 (2013)

# List of Figure Captions

**Figure 1** Schematic of the ground plane cloak.

**Figure 2** Simulated temperature profile for proposed carpet device. (a) Thermal cloak for heat flux flows from top side to bottom. (b) Thermal cloak for heat flux flows from left side to right. The isothermal lines are plotted by using white color.

**Figure 3** Schematic illustration of the experimental setup and Photo of the fabricated sample (a) Schematic illustration of the carpet cloak experiment. (b) Photo of the fabricated thermal carpet device. The composite is assembled by bulk stainless steel with thermal conductivity 16.2 W m$^{-1}$ K$^{-1}$ and cork with thermal conductivity 0.065 W m$^{-1}$ K$^{-1}$.

**Figure 4** Measured temperature distribution for steady state. (a) Temperature distribution when the top and bottom sides are set to be 27.3$^{o}$C and 9.1$^{o}$C. (b) Temperature distribution when the top and bottom sides are set to be 43.4$^{o}$C and 6.5$^{o}$C. (c) Temperature distribution when the left and right sides are set to be 25.1$^{o}$C and 8.6$^{o}$C. (d) Temperature distribution when the left and right sides are set to be 37.9$^{o}$C and 9.9$^{o}$C.

**Figures**

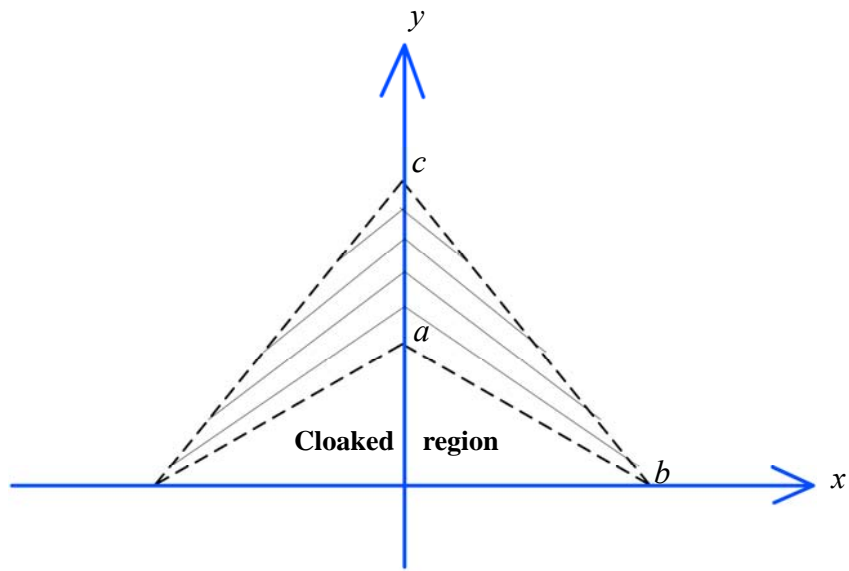

Figure 1

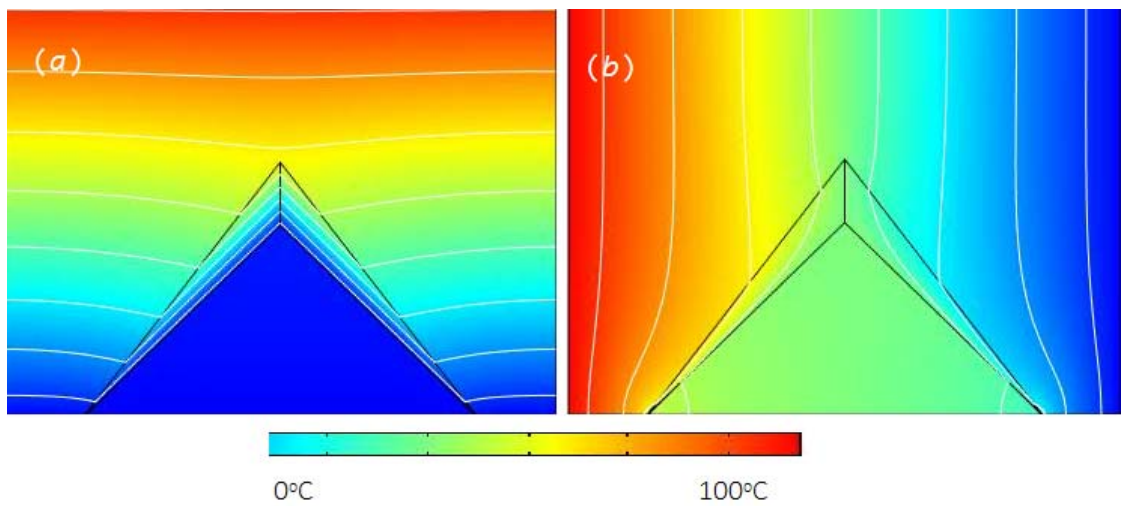

Figure 2.

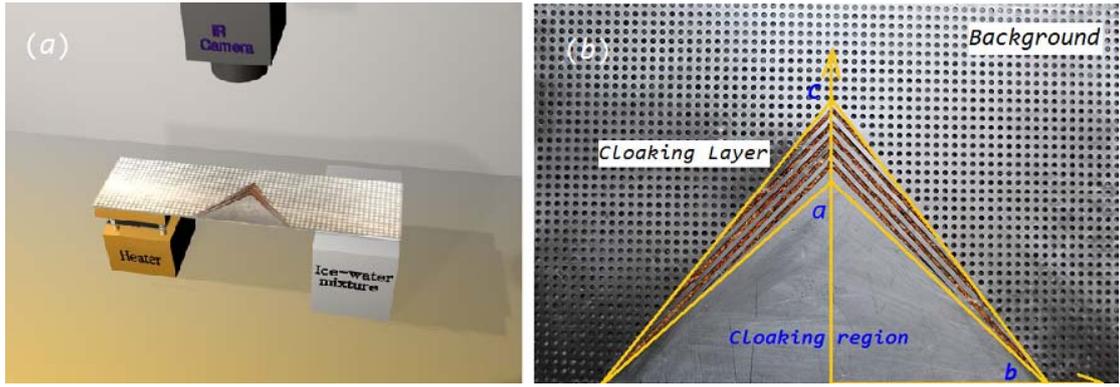

Figure 3

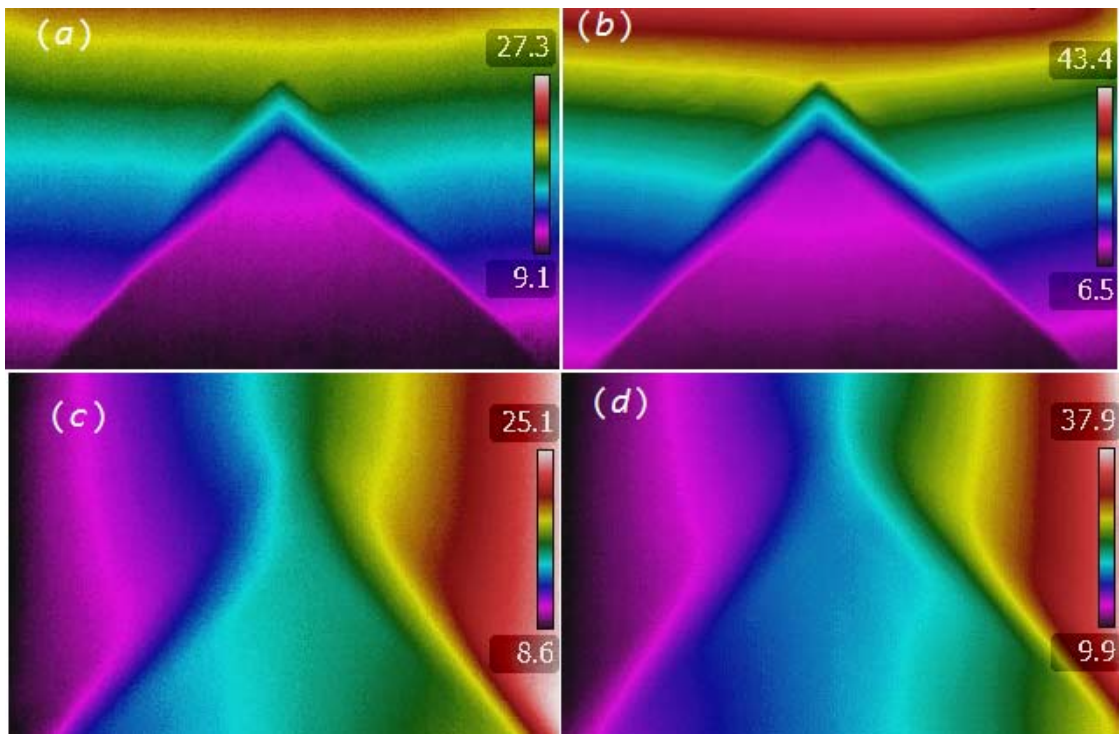

Figure 4